\documentclass[preprint]{aastex}
\def\simgr{\,\hbox{\hbox{$ > $}\kern -0.8em \lower 1.0ex\hbox{$\sim$}}\,}
\def\simle{\,\hbox{\hbox{$ < $}\kern -0.8em \lower 1.0ex\hbox{$\sim$}}\,}

\shortauthors{THORSTENSEN}
\shorttitle{Evolved CV Secondary}

\def\css1340{CSS J134052.0+151341} 

\begin{document}
\title{\css1340 : A Cataclysmic Binary Star with a Stripped, Evolved Secondary
\footnote{Based on observations obtained at the MDM Observatory, operated by
Dartmouth College, Columbia University, Ohio State University, 
Ohio University, and the University of Michigan.}
}

\author{John R. Thorstensen}
\affil{Department of Physics and Astronomy\\
Dartmouth College\\
Hanover NH, 03755\\
}

\begin{abstract}
I report spectroscopy and time-series photometry of the cataclysmic
binary \css1340.  The optical light is dominated by the secondary
star, which I classify as K4 ($\pm 2$ subclasses), yet the orbital
period derived from the absorption radial velocities is only
2.45 hr, implying a Roche radius much too small to contain
a main-sequence K star.  The spectrum shows enhanced sodium
absorption in several lines, suggesting that the surface material
has been processed at high temperatures.  \css1340 appears to be
a rare example of a cataclysmic binary in which the secondary
star is the stripped core of a formerly much more massive star,
that began mass transfer after much of the core's nuclear
evolution had taken place.
\end{abstract}

\keywords{keywords: stars}

\section{Introduction}

Cataclysmic variable stars (CVs) are close binary systems in which
a white dwarf accretes matter from a less-compact companion that
fills its Roche critical lobe.  The companion usually resembles
a low-mass main sequence star.  \citet{faulkner72} pointed
out that if a star fills its Roche critical lobe in a close
binary, its mean density $\left<\rho\right>$ is inversely related
to the orbital period $P_{\rm orb}$, 
with little dependence on the masses of the two stars.  Along 
the main sequence, the mean density increases strongly with
decreasing mass.  Consequently, the secondary stars in short-period
CVs tend to resemble late M dwarfs.  At periods 
shorter than $\sim$ 4 hr, 
the secondary is usually not detectable in the combined
spectrum, because other sources of 
light overwhelm its feeble contribution.
At longer periods, the secondaries tend to be
hotter, more luminous, and more conspicuous.  The relationship
between $P_{\rm orb}$ and the secondary spectral type $SpT$ 
is explored in detail by \citet{kniggedonor}.

Some CVs depart radically from the typical $SpT$-$P_{\rm orb}$ 
relationship.  If the secondary star is also a degenerate
dwarf, the system is called an AM CVn star; these can have
$P_{\rm orb} < 10$ min.  Over a dozen examples
are known \citep{nelemans05, roelofs07, rau10}.  

Here we are concerned with another CV subtype -- 
the small number of short-period systems in which the 
secondary star not a white dwarf, but is significantly warmer than 
expected for the orbital period.
EI Psc (= 1RXS J232953.9+062814) has $P_{\rm orb}$ = 64 min,
yet shows a conspicuous K-type secondary \citep{thoreipsc}.
The orbital period of QZ Ser is less extreme at 2.0 hours,
but it too has a K-type secondary \citep{thorqzser}.  
\citet{littlefair06} obtained time-series photometry of
the 2.4-hour eclipsing system
SDSS J170213.26+322954.1 and derived a temperature 
for the secondary consistent with an M0 star, while
the typical relation would predict $\sim$ M4. \citet{szkodysdssiii}
found a spectral type of M1.5 $\pm$ 1.1 for the
secondary, marginally consistent with the \citet{littlefair06}
result, and also hotter than expected for $P_{\rm orb}$.

The standard explanation for these warmer-than-expected
secondaries is that mass transfer began after the
secondary had consumed most of the hydrogen in the its
core.  This greatly alters the mass-luminosity-temperature
($MLT$) relation, and in particular makes $\left<\rho\right>$
greater at a given surface temperature, which results in 
the departure from the usual $SpT$-$P_{\rm orb}$
relation.  This idea is supported by evolutionary
calculations \citep{thoreipsc} and by indications of 
unusual abundances in the accreted material \citep{gaensicke03}
and in the secondary's photosphere.  In particular,
\citet{thorqzser} found unusually strong sodium
absorption lines in the QZ Ser secondary spectrum, which
they interpreted as evidence that the material had
been processed at a temperature high enough to
drive proton capture by $^{22}$Ne.
One attractive feature of the evolved-secondary
scenario is that, if
the secondary were more massive than the
primary when it first contacted its Roche critical lobe, 
the subsequent mass transfer would have been 
unstable, leading to rapid shrinkage of the orbit.  
This could explain to the short periods observed
\citep{littlefair06}.

A stripped-core secondary also features in at least
one neutron-star binary, namely AY Sex 
\citep[FIRSTJ102347+003841; ][]{archibald09,ta05}.
The primary in this system is a 500 Hz millisecond
pulsar, and the secondary is a mid-G
star.  A recent VLA parallax measurement by
\citet{deller12}, combined with the constraints derived
by \citet{ta05}, give a secondary-star mass of only
0.24 M$_\odot$, so it is {\it extremely} undermassive
for its spectral type.  AY Sex also shows
some evidence of Na enhancement \citep{ta05}.

Here I report on a new member of this exclusive class,
\css1340.  This was discovered in the Catalina Real Time
Survey \citep[CRTS; ][]{drakecrts}, which has
turned up a very large number of new CV candidates
\citep{ww12, thorcrts} . 
Table \ref{tab:properties} summarizes some
pertinent information about the star.
The CRTS light curve of this object
\footnote{This and all other CRTS information
about this star were
retrieved the
on-line CRTS CV catalog, available at 
http://nesssi.cacr.caltech.edu/catalina/AllCV.html}
shows with a white-light magnitude
just fainter than 18th, and a single outburst to 14.5
magnitude starting 2010 May 31.  While the outburst behavior
is unremarkable, the magnitude at minimum is
unusually steady, varying by only a few tenths
of a magnitude at most.  Most dwarf novae
show much larger variations at minimum.

\section{Observations}

All the observations are from MDM Observatory, on 
Kitt Peak, Arizona.  Spectra were obtained on 2013
Mar.~3, 4, and 5 UT; on the 5th, I also obtained
simultaneous time-series photometry. 

\subsection{Spectroscopy}

The spectra are from the 2.4m Hiltner telescope
and modular spectrograph, equipped with a 600 line mm$^{-}$
grating and the SITe 1024$^2$ CCD detector `Templeton'.  
The region from 4660 to 6732 \AA\ was covered at 2 \AA\ 
pixel$^{-1}$, and the 1.1-arcsec slit gave 
resolution somewhat better than 2 pixels FWHM.  
Standard stars were observed in twilight when the 
weather was clear, but unfortunately many of the
spectra of the program object were taken through thin
cloud or in poor seeing, so not all are of good 
photometric quality.  Exposures of wavelength
calibration lamps taken in twilight gave the 
base pixel-to-wavelength relation, which typically
had rms residuals $\sim 0.03$ \AA , and the reduction 
pipeline shifted 
this solution to force the [OI] $\lambda 5577$ line
(present in all the spectra) to its rest 
wavelength. Exposure times were mostly 720 s
to suppress readout noise and minimize dead time.

The mean spectrum (Fig.~\ref{fig:specs}, top panel)
shows the absorption lines of a late-type star
and weak emission at H$\alpha$.  I estimated the
spectral type and light fraction of the secondary
by scaling and subtracting spectra of stars 
classified by \citet{keenan89} and looking
for the best cancellation of late-type features.
This constrained the spectral type to be 
K4 $\pm$ 2 subclasses.  The secondary contributes
most -- possibly almost all -- of the 
system's light.

Along with the observed spectrum, Fig.~\ref{fig:specs} 
shows the result of subtracting a scaled K4 spectrum.
The subtracted spectrum shows stronger H$\alpha$
emission than the original, because even K stars
have fairly strong H$\alpha$ absorption, which 
masks the intrinsically weak emission line.  
Also, NaD is grossly undersubtracted.  Comparison
with spectral type standards shows that the 
NaD absorption in the original spectrum is 
roughly as strong as that in an M0 star.  The
sodium blend near $\lambda$5685  
is also badly undersubtracted,
and the Na blend near $\lambda$6158 may also be enhanced,
though in the latter case the strong CaI lines
near $\lambda$6160 complicate the interpretation.  A
similar apparent Na enhancement is seen in 
the secondary of QZ Ser \citep{thorqzser}.

\begin{figure}
\plotone{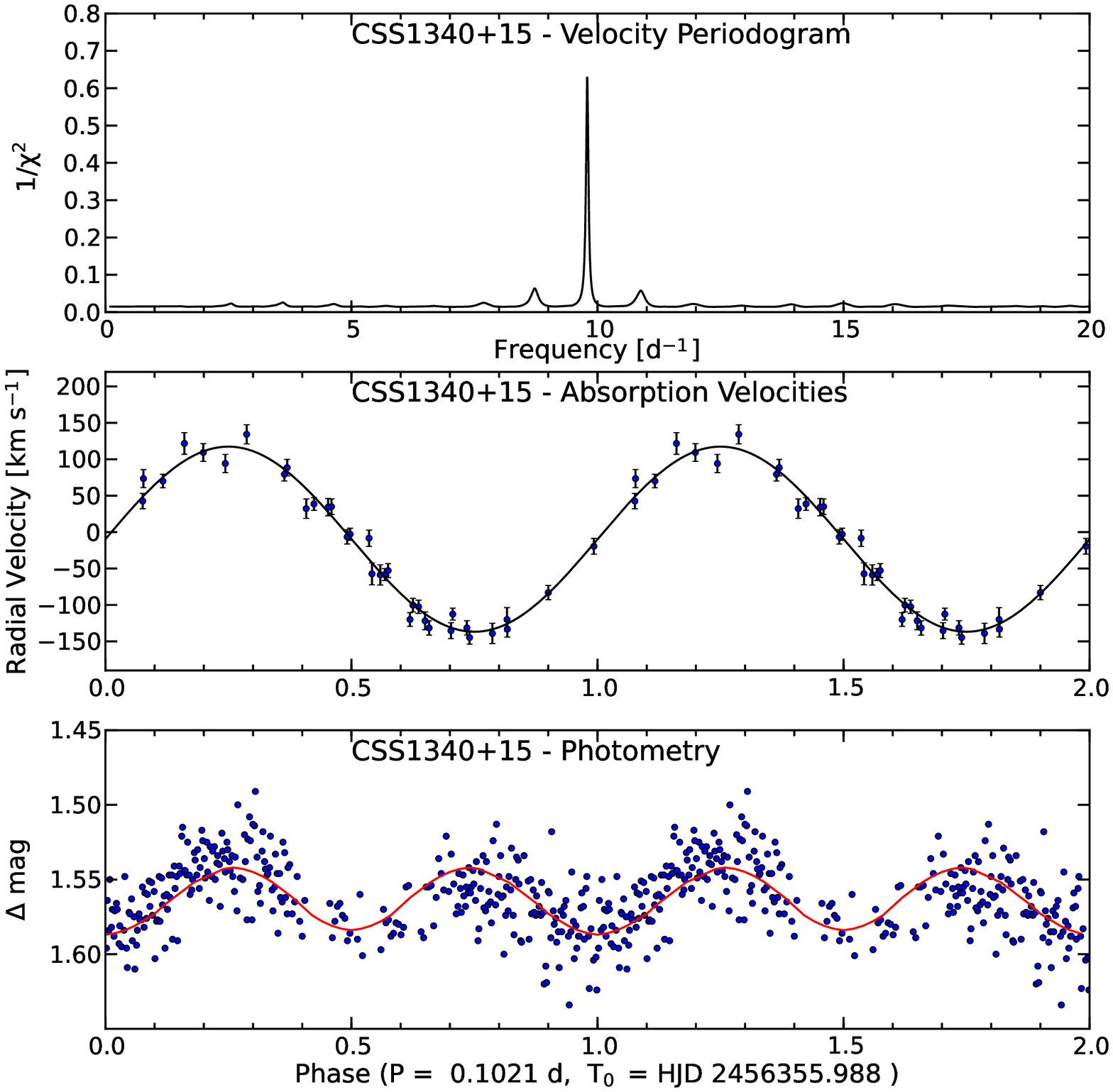}
\caption{{\it Upper panel:} Periodogram of the absorption-line
velocities. 
{\it Middle panel:} Absorption-line velocities folded on the 
orbital period, with the best-fitting sinusoid superposed.
{\it Lower panel:} White-light magnitudes, folded on the 
same ephemeris as the middle panel.  The solid curve
is derived from the ellipsoidal variation model described in the
text.
}
\label{fig:montage2}
\end{figure}

The H$\alpha$ line was too weak for radial
velocity measurements, but the rich absorption
spectrum yielded usable velocities for 34 of the
49 spectra (Table \ref{tab:velocities}).  
These were measured using 
the cross-correlation routine {\it xcsao} in the 
{\it rvsao} package \citep{kurtzmink}; 
I used a velocity-compensated
composite of G- and K-type velocity standards as
a template. 
Fig.~\ref{fig:montage2} shows the result of a 
period search of the velocities, using the
sine-fitting method described by \citet{tpst};
the orbital period is obvious.
Fitting a sinusoid of the form
$$v(t) = \gamma + K \sin (2 \pi (t - T_0) / P)$$
yields 
\begin{eqnarray} 
T_0 &=& \hbox{HJD } 2456355.9877 \pm 0.0007, \nonumber \\
P &=& 0.10213 \pm 0.00008\ {\rm d},\nonumber \\
K &=& 127 \pm 5\ {\rm km\ s^{-1}},\nonumber \\
\hbox{and} && \nonumber \\
\gamma &=& -10 \pm 4\ {\rm km\ s^{-1}}.\nonumber 
\end{eqnarray}
The middle panel of Fig.~\ref{fig:montage2} shows
the cross-correlation velocities folded on this
period, together with the best-fit sinusoid, and 
the lower panel of Fig.~\ref{fig:specs} shows
a greyscale representation of the data in the 
vicinity of the NaD line.  

\subsection{Time-series Photometry}

On 2013 March 5 UT, I obtained simultaneous
time-series photometry with the MDM 1.3m McGraw-Hill
telescope and an Andor Ikon DU-937N camera.
The Andor camera's sensor is a thinned, backside-illuminated,
frame-transfer CCD; the active area is has 512 $\times$ 512,
pixels, each 13 $\mu$m square.  Binning
this $4 \times 4$ gave a 128$^2$ data array
subtending 139 arcsec on a side.  Exposures
were 45 seconds through a Schott GG420 filter,
chosen to suppress scattered moonlight.  The 
vendor-supplied {\it Solis} software wrote the images
as a FITS data cube, which I reduced using
Python-languages programs based on PyFits
\citep{pyfits} and NumPy.  
Processing steps included subtracting bias and
dark images, and dividing by a flat field
derived from images of the twilight sky.

The program star and several others were measured
in each frame using the aperture photometry 
task {\it phot} from the IRAF implementation of
DAOphot.  Frames were centered southwest of the 
program object in order to include a comparison
star 102 arcsec to the west and 89 arcsec south,
which is $\sim 1.5$ mag brighter.  Intermittent
cirrus clouds greatly degraded the signal-to-noise
of some of the images, so images in which the 
comparison star was attenuated by more than $\sim 1$ mag
were discarded from the analysis.  

The lower panel of Fig.~\ref{fig:montage2} shows
the magnitudes relative to the comparison star,
folded on the orbital ephemeris.  A clear
double-peaked modulation is present.
Comparison with the velocity curve shows that the fainter
minimum coincides with inferior conjunction of the 
secondary, as expected for ellipsoidal variation.  
This modulation accounts for all the variation -- 
the scatter around the ellipsoidal light-curve 
is comparable to the scatter in the measurements
of similarly faint stars in the field.

\section{Discussion}

At $P_{\rm orb} = 2.45$ hr, the $SpT$-$P_{\rm orb}$ 
relation for normal CVs predicts a secondary 
around M3 or M4 \citep{kniggedonor}.  At K4 $\pm$ 2,
the secondary is clearly well to the warm side
of the curve.  The 
apparent Na enhancement reinforces the resemblance
to QZ Ser.  \css1340 joins the exclusive 
club of stripped-core secondaries.  

\begin{figure}
\plotone{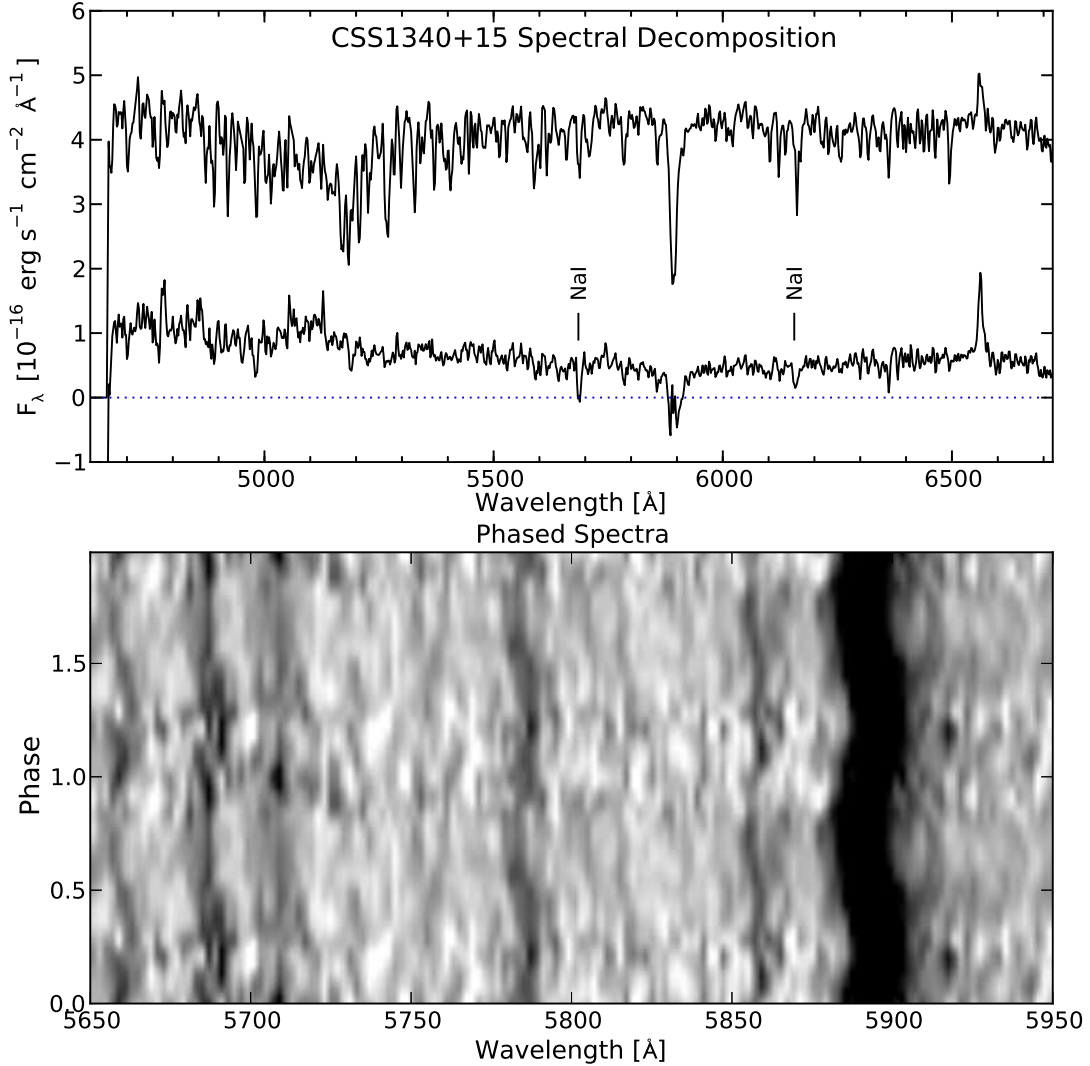}
\caption{{\it Upper panel:} Mean spectrum. The individual spectra were shifted
to zero velocity using the absorption-line ephemeris prior to 
averaging.  The lower trace shows the residual after subtracting
a scaled spectrum of the K4 star Gliese 570a.  Note the 
poor cancellation of the NaD lines, as well as the weaker
Na features marked, and how correction for the K-star's
H$\alpha$ absorption ($\lambda$6563) enhances visibility of the
emission line.  The dotted horizontal line indicates the zero level.
{\it Lower panel:} A greyscale representation of the spectra in the 
vicinity of the NaD lines. Each line in the image is derived
from a running average of the rectified spectra that lie near the
nominal orbital phase.  The orbital motion is apparent, and the 
enhanced Na feature near $\lambda$5685 is also visible.
}
\label{fig:specs}
\end{figure}

The amplitude of the ellipsoidal variation is 
fairly small, so the inclination is fairly close
to face-on.  To quantify this, I adapted the 
computer code developed by \citet{ta05} 
(to model the AY Sex light curve) to the present case.
The icosahedral tesselation of the secondary
star, surface-brightness/color/effective-temperature
relations, and prescriptions
for limb- and gravity-darkening were unchanged.
I adopted $T_{\rm eff} = 4345$ kelvin 
as appropriate for a K4V photosphere \citep{pickles}. 
The primary star's luminosity (which heats
one face of the AY Sex secondary) was turned
off, but to account for a possible accretion
disk contribution I added extra light comprising 
about 15 per cent of the system's light, guided
by the spectral decomposition in Fig.~\ref{fig:specs}
(top panel).  In the original case of AY Sex,
$B$, $V$, and $I$ bands light curves were available, 
but the present data are in 
white light, so the color-fitting portions of the 
code were turned off.  I assumed that the
white-light behavior is represented adequately 
by a $V$-band model, which seems reasonable given the
available signal-to-noise.  I assumed
further that the light curve oscillates 
around $V = 18.3$, from the data in Table
\ref{tab:properties}.  

A model with $M_1 = 0.7$ M$_{\odot}$, $M2 = 0.5$ M$_{\odot}$,
and an inclination $i = 27$ degrees gave
excellent agreement, both with the light curve
and the observed velocity amplitude.  
Assuming a reddening
$E(B-V) = 0.034$ \citep{schlegel98}, the
$V = 18.3$ normalization puts this model at 790 pc.  
This distance will scale approximately as
the secondary mass to the 1/3 power, and 
inversely with the surface brightness, which 
increases steeply with the assumed effective
temperature. 
This model is not intended to be unique,
but rather to check whether the available
information supports a primary star
mass in the white dwarf range, which it
evidently does.

It is interesting that only one 
outburst has been observed.  EI Psc and 
QZ Ser also outburst rarely.  
In the absence of an outburst, an object
similar to \css1340 would be inconspicuous.
This raises
the question of how many
similar objects may remain undetected.
The optical spectrum is dominated by a K star, 
with weak emission, so emission line surveys
such as IPHAS \citep{withamiphas} would not see it.
Color-based surveys would fare no better;
the optical colors (Table~\ref{tab:properties}) are 
typical of K stars \citep{covey07}, and
even the $u-g$ color does not stick out.  
It seems likely, then, that a sizeable
number of these objects remain undetected.
This may change in the future; 
the Large Synoptic Survey Telescope 
should be capable of detecting 
quiescent objects of this kind
through their ellipsoidal variation -- most
will be closer to edge-on than this one -- or
through eclipses.  The GAIA astrometric
satellite may also contribute, since the secondary
stars in similar objects are likely to 
have radii significantly smaller than 
normal stars of the same color, leading
to distances that are anomalously 
small for the color and apparent magnitude.  

It also may be significant that this object
has a period squarely in the middle 
of the so-called `period gap' \citep[and references therein]{kolbstillthere}, the
interval $2\ {\rm hr} < P_{\rm orb} < 3\ {\rm hr}$
where relatively few CVs (especially non-magnetic
dwarf novae) are found.  Interestingly,
SDSS J170213.26+322954.1 , the warm-secondary eclipsing 
system noted in the 
Introduction, is also well within the gap
at almost the same $P_{\rm orb}$, 2.4 hr
\citet{littlefair06} .

A popular explanation for 
the gap, the {\it disrupted
magnetic braking} scenario, is as follows
\citep[and references therein]{kniggedonor}. 
CVs evolve to shorter periods by angular
momentum loss, their secondaries become
distended because they go out of
thermal equilibrium as mass is removed
from their surfaces.  Around 3 hr, an abrupt change
in the angular momentum loss 
(usually thought to involve a 
reconfiguration of the secondary's magnetic
field) quenches the angular momentum 
loss, causing the secondary to relax
back toward its natural radius and 
detach from the Roche lobe. The system
evolves slowly across the gap due to 
more gradual angular momentum loss, and
lights up again around 2 hr as the 
Roche lobe squeezes down onto the 
secondary.   

\css1340 is relatively quiescent, but it
is not entirely so -- it shows H$\alpha$
emission and erupted in the recent past.
Unless mass transfer is being driven through
some mechanism other than Roche overflow, 
It must be very close to its Roche lobe. 
Disrupted magnetic braking would require
that a CV in the middle of the gap be
well-detached from its Roche lobe, so
this system must have evolved differently.
As \citet{littlefair06} suggest, it is possible
that the system was dropped into 
nearly its present configuration in a sudden 
episode of mass loss driven by dynamical-timescale 
mass transfer.

\acknowledgments
I gratefully acknowledge support from NSF grant
and AST-1008217, and thank the MDM Observatory staff for 
their cheerful and excellent support.

\begin{deluxetable}{lll}
\tablecolumns{3}
\tablewidth{0pt}
\tablecaption{Properties of \css1340}
\tablehead{
\colhead{Property} &
\colhead{Value} &
\colhead{Reference\tablenotemark{a}} \\
}
\startdata
$\alpha_{2000}$ & 13$^{\rm h}$ 40$^{\rm m}$ 52$^{\rm s}$.062  & PPMXL \\
$\delta_{2000}$ & +15$^\circ 13' 40".79$ & \\[1.2ex]
Outburst date & 2010 May 31 & CRTS \\
Outburst magnitude & 14.48 & \\[1.2ex]
u & $20.259 \pm 0.042$ & SDSS \\
g & $18.731 \pm 0.008$ & \\
r & $17.988 \pm 0.007$ & \\
i & $17.721 \pm 0.007$ & \\
z & $17.572 \pm 0.015$ & \\[1.2ex]
V & 18.30\tablenotemark{b}  &  \\[1.2ex]
Y & $16.865 \pm 0.011$ & UKDISS \\
J & $16.528 \pm 0.011$ &  \\
H & $16.122 \pm 0.014$ &  \\
K & $15.926 \pm 0.020$ & \\[1.2ex]
\enddata
\tablenotetext{a}{References are as follows: PPMXL is the PPMXL
astrometric catalog \citep{ppmxl}; CRTS is the Catalina
Real Time Survey \citep{drakecrts}; 
UKDISS is the UK Deep Infrared Sky Survey \citep{lawrence07, hambly08},
Data Release 8;
SDSS is the SDSS Photometric Catalog, Data Release 9. 
} 
\tablenotetext{b}{The $V$ magnitude is computed using 
$$V = g - 0.5784 (g - r) - 0.0038,$$
an approximation derived by R. H. Lupton and cited at the
SDSS website.
}
\label{tab:properties}
\end{deluxetable}

\begin{deluxetable}{lrr}
\tablecolumns{3}
\tablewidth{0pt}
\tabletypesize{\scriptsize}
\tablecaption{Radial Velocities}
\tablehead{
\colhead{Time} &
\colhead{$v_{\rm rad}$} &
\colhead{$\sigma$} \\
\colhead{} &
\colhead{[km s$^{-1}$]} &
\colhead{[km s$^{-1}$]}
}
\startdata
6355.0097  & $   39$ & $   9$  \\
6355.0172  & $   -3$ & $   8$  \\
6355.0244  & $  -58$ & $   8$  \\
6355.0314  & $ -103$ & $   9$  \\
6355.0385  & $ -113$ & $   8$  \\
6355.8405  & $  -59$ & $  14$  \\
6355.8498  & $ -122$ & $  12$  \\
6355.8668  & $ -120$ & $  16$  \\
6355.9325  & $   35$ & $  11$  \\
6355.9409  & $  -57$ & $  15$  \\
6355.9494  & $ -100$ & $   9$  \\
6355.9606  & $ -131$ & $  10$  \\
6355.9690  & $ -133$ & $  11$  \\
6355.9775  & $  -83$ & $  10$  \\
6355.9956  & $   74$ & $  12$  \\
6356.0041  & $  122$ & $  15$  \\
6356.0126  & $   94$ & $  13$  \\
6356.7443  & $   32$ & $  13$  \\
6356.7528  & $   -6$ & $  10$  \\
6356.7613  & $  -53$ & $  10$  \\
6356.7698  & $ -132$ & $  10$  \\
6356.7782  & $ -145$ & $   9$  \\
6356.8040  & $  -19$ & $  11$  \\
6356.8125  & $   43$ & $  11$  \\
6356.8419  & $   79$ & $   9$  \\
6356.9188  & $   70$ & $   9$  \\
6356.9272  & $  109$ & $  12$  \\
6356.9362  & $  134$ & $  13$  \\
6356.9446  & $   89$ & $  11$  \\
6356.9531  & $   34$ & $  12$  \\
6356.9616  & $   -8$ & $  11$  \\
6356.9701  & $ -120$ & $   9$  \\
6356.9786  & $ -135$ & $  11$  \\
6356.9872  & $ -139$ & $  14$  \\
\enddata
\tablecomments{Absorption-line radial
velocities.  Times listed are the heliocentric
Julian date of mid-integration, minus 2 450 000.
The time base is UTC.
}
\label{tab:velocities}
\end{deluxetable}

\begin{thebibliography}

\bibitem[Archibald et al.(2009)]{archibald09} Archibald, A.~M., 
Stairs, I.~H., Ransom, S.~M., et al.\ 2009, Science, 324, 1411 

\bibitem[Barrett \& Bridgman(2000)]{pyfits} Barrett, P.~E., \& 
Bridgman, W.~T.\ 2000, in 
Astronomical Society of the Pacific Conference Proceedings, vol 216, 
Astronomical Data Analysis Software and Systems IX, ed. N. Manset, 
C. Veillet, \& D. Crabtree , 67 

\bibitem[Covey et al.(2007)]{covey07} Covey, K.~R., Ivezi{\'c}, 
{\v Z}., Schlegel, D., et al.\ 2007, \aj, 134, 2398 

\bibitem[Deller et al.(2012)]{deller12} Deller, A.~T., 
Archibald, A.~M., Brisken, W.~F., et al.\ 2012, \apjl, 756, L25 

\bibitem[Drake et al.(2009)]{drakecrts} Drake, A.~J., et al.\
2009, \apj, 696, 870 

\bibitem[Faulkner et al.(1972)]{faulkner72} Faulkner, J., 
Flannery, B.~P., \& Warner, B.\ 1972, \apjl, 175, L79 

\bibitem[G{\"a}nsicke et al.(2003)]{gaensicke03} G{\"a}nsicke, 
B.~T., Szkody, P., de Martino, D., et al.\ 2003, \apj, 594, 443 

\bibitem[Hambly et al.(2008)]{hambly08} Hambly, N.~C., Collins, 
R.~S., Cross, N.~J.~G., et al.\ 2008, \mnras, 384, 637 

\bibitem[Keenan \& McNeil(1989)]{keenan89}
Keenan, P.~C., \& McNeil, R.~C.\ 1989, \apjs, 71, 245

\bibitem[Lawrence et al.(2007)]{lawrence07} Lawrence, A., Warren, 
S.~J., Almaini, O., et al.\ 2007, \mnras, 379, 1599 

\bibitem[Knigge(2006)]{kniggedonor} Knigge, C.\ 2006, \mnras, 373,
484

\bibitem[Kolb et al.(1998)]{kolbstillthere} Kolb, U., King, A.~R.,
\& Ritter, H.\ 1998, \mnras, 298, L29

\bibitem[Kurtz \& Mink(1998)]{kurtzmink} Kurtz, M.~J. \& Mink,
D.~J.\ 1998, \pasp, 110, 934

\bibitem[Littlefair et al.(2006)]{littlefair06} Littlefair, S.~P., 
Dhillon, V.~S., Marsh, T.~R.,  
G{\"a}nsicke, B.~T.\ 2006, \mnras, 371, 1435 

\bibitem[Nelemans(2005)]{nelemans05} Nelemans, G.\ 2005, in 
Astronomical Society of the Pacific Conference Series, vol. 330,
The Astrophysics of Cataclysmic Variables and Related Objects, 
ed. J.-M. Hameury \& J.-P. Lasota, 27 

\bibitem[Pickles(1998)]{pickles} Pickles, A.~J.\ 1998, \pasp, 
110, 863 

\bibitem[Roeser et al.(2010)]{ppmxl} Roeser, S., Demleitner,
M., \& Schilbach, E.\ 2010, \aj, 139, 2440

\bibitem[Rau et al.(2010)]{rau10} Rau, A., Roelofs, G.~H.~A., 
Groot, P.~J., et al.\ 2010, \apj, 708, 456 

\bibitem[Roelofs et al.(2007)]{roelofs07} Roelofs, G.~H.~A., 
Nelemans, G., \& Groot, P.~J.\ 2007, \mnras, 382, 685 

\bibitem[Schlegel, Finkbeiner, \& Davis(1998)]{schlegel98}
Schlegel, D. J., Finkbeiner, D. P., \& Davis, M. 1998, \apj, 500, 525

\bibitem[Szkody et al.(2004)]{szkodysdssiii} Szkody, P., Henden, A., 
Fraser, O., et al.\ 2004, \aj, 128, 1882 

\bibitem[Thorstensen \& Armstrong(2005)]{ta05} Thorstensen, J.~R., \& Armstrong, E.\ 2005, \aj, 130, 759 

\bibitem[Thorstensen et al.(2002a)]{thoreipsc} Thorstensen, J.~R., 
Fenton, W.~H., Patterson, J.~O., et al.\ 2002a, \apjl, 567, L49 

\bibitem[Thorstensen et al.(2002b)]{thorqzser} Thorstensen, J.~R., 
Fenton, W.~H., Patterson, J., et al.\ 2002b, \pasp, 114, 1117 

\bibitem[Thorstensen et al.(1996)]{tpst} Thorstensen, J. R., Patterson, J.,
Thomas, G., \& Shambrook, A. 1996, \pasp, 108, 73

\bibitem[Thorstensen \& Skinner(2012)]{thorcrts} Thorstensen, J.~R., \& Skinner, J.~N.\ 2012, \aj, 144, 81 

\bibitem[Witham et al.(2007)]{withamiphas} Witham, A.~R., Knigge,
C., Aungwerojwit, A., et al.\ 2007, \mnras, 382, 1158

\bibitem[Woudt et al.(2012)]{ww12} Woudt, P.~A., Warner, B.,
de Bud{\'e}, D., et al.\ 2012, \mnras, 2533


\end{thebibliography}
\end{document}